\documentclass{aastex}
\usepackage{spr-astr-addons}
\usepackage{url}
\RequirePackage{color}

\newcommand{\emaila}{a78b@yandex.ru}

\begin{document}

\title{Timing of eclipsing binary V0873 Per: a third body candidate}
\slugcomment{Not to appear in Nonlearned J., 45.}
\shorttitle{V0873 Per: a third body candidate}
\shortauthors{Bogomazov et al.}

\author{A. I. Bogomazov\altaffilmark{1}}
\email{\emaila}
\and
\author{M. A. Ibrahimov\altaffilmark{2}}
\and
\author{B. L. Satovsky\altaffilmark{3}}
\and
\author{V. S. Kozyreva\altaffilmark{1}}
\and
\author{T. R. Irsmambetova\altaffilmark{1}}
\and
\author{V. N. Krushevska\altaffilmark{4}}
\and
\author{Yu. G. Kuznyetsova\altaffilmark{4}}
\and
\author{E. R. Gaynullina\altaffilmark{5}}
\and
\author{R. G. Karimov\altaffilmark{5}}
\and
\author{Sh. A. Ehgamberdiev\altaffilmark{5}}
\and
\author{A. V. Tutukov\altaffilmark{2}}
\and

\altaffiltext{1}{M. V. Lomonosov Moscow State University, Sternberg Astronomical Institute, 13, Universitetskij prospect, Moscow, 119991, Russia}
\altaffiltext{2}{Institute of astronomy, Russian Academy of Sciences, 48, Pyatnitskaya ulitsa, Moscow, 119017, Russia}
\altaffiltext{3}{AstroTel Ltd., 1A, Nizhegorodskaya ulitsa, Moscow, 109147, Russia}
\altaffiltext{4}{Main Astronomical Observatory, National Academy of Sciences
of Ukraine, 27, Akademika Zabolotnoho ulitsa, Kyiv, 03680, Ukraine}
\altaffiltext{5}{Ulugh Beg Astronomical Institute, Uzbek Academy of
Sciences, 33, Astronomicheskaya ulitsa, Tashkent, 100052,
Uzbekistan}

\begin{abstract}
We analyse a set of moments of minima of eclipsing variable V0873
Per. V0873 Per is a short period low mass binary star. Data about moments of minima of V0873 Per were
taken from literature and our observations during
2013-2014. Our aim is to test the system on existence of new
bodies using timing of minima of eclipses. We found the periodical
variation of orbital period of V0873 Per. This variation can be explained
by the gravitational influence of a third companion on the central
binary star. The mass of third body candidate is $\approx 0.2 M_{\odot}$, its
orbital period is $\approx 300$ days. The paper also includes a
table with moments of minima calculated from our observations
which can be used in future investigations of V0873 Per.
\end{abstract}


\section{Introduction}

Eclipsing binary V0873 Per (TYC 2853-18-1) was discovered by
\citet{nicholson2006} using
published data from the Northern Sky Variability Survey
\citep{wozniak2004}\footnote{See also General Catalogue of Variable Stars \citep{samus2014}.}. The V magnitude range of this system is $10.8^m-11.5^m$ and its light curve was identified as an EW-Type, orbital period was found to be $\approx 0.2949$ days. \citet{nicholson2006} give the next ephemeris for V0873 Per:

\begin{equation}
\label{eph1}
Min\ I= HJD 2451370.87525 + 0.2949\times E;
\end{equation}

\citet{samec2009} conducted spectral and photometry investigations of V0873 Per. They estimated spectral class of the primary (more massive) star as K0V$\pm$1 (effective temperature $T_{eff}=5150\pm 150 K$), while spectral class of secondary component was not studied. V0873 Per was found to be among the W-type W UMa contact binaries. \citet{samec2009} solution reveals that the less massive, slightly hotter component is eclipsed at phase zero. The mass ratio of the system is $M_1/M_2 \approx 2.6$ and is likely tending toward larger mass ratios as the secondary component is consumed by the primary. \citet{samec2009} obtained their own ephemeris for V0873 Per:

\begin{eqnarray}
\label{eph2}
Min\ I= HJD 2451370.875 + 0.2949039\times E;
\end{eqnarray}

\citet{kriwattanawong2015} presented observations of V0873 Per in {\it BVR} filter bands. Photometric solutions showed that this system is a W-type with a mass ratio of q = 2.504 ($\pm 0.0029$), confirming the results of \citet{samec2009}. The derived contact degree was found to be f = 18.10\%($\pm 1.36$\%). \citet{kriwattanawong2015} suspected the cyclic variation with the period of about 4 years that could be due to existence of the third companion in the system or the mechanism of magnetic activity cycle in the binary.

The aim of this paper is to study V0873 Per using timing of light minima from literature and our observations in order to find possibly existing new bodies in this system. We found a possible evidence in favour of existence of a third body with a mass equal to mass of a tiny star.

\section{Observations and data reduction}

\begin{figure}
\includegraphics{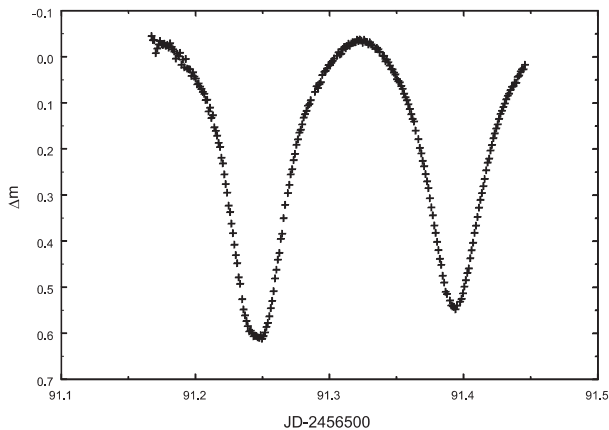}
\vspace{0pt} \caption{Sample light curve of V0873 Per in {\it R$_c$} filter
during our observations at Zeiss-600 telescope in Crimea.}\label{f1}
\end{figure}

Photometric observations of eclipsing binary system V0873 Per were performed during September, October and November 2013 (10 nights) and August 2014 (4 nights) at South Station of M. V. Lomonosov Moscow State University in Nauchnij, Crimea at the altitude about 600 meters above sea level. We used an Apogee CCD camera Ap-47p ($1024\times 1024$ pixels) attached to the Cassegrain focus of a Zeiss-600 telescope. Computer clock was synchronized using GPS signals every evening before and during the observations.

Observational material was obtained as a result of continuous and prolonged (6-8 hours) monitoring in {\it R$_c$} band. The exposure time was set to be 25-30 seconds, therefore typical measurement errors were
$\approx 0.002^{m}-0.003^{m}$. Exposure and readout cycle take about 1.5 minutes during observations. During some observational nights we used a field lens in front of the CCD camera to increase
the field of view from $6\times 6$ arc min to $13\times 13$ arc min. In Figure \ref{f1} we present a sample light curve of V0873 Per in {\it R$_c$} band during our observations at South Station.

Also photometric observations of eclipsing binary system V0873 Per
were performed at Maidanak observatory of Ulugh Beg Astronomical
Institute of Uzbek Academy of Sciences using 60 cm Zeiss-600
(North) telescope with a CCD camera FLI PL 1K$\times$1K. During 20
observational nights in September, October and November 2013
and 5 observational nights in August and September 2014 we obtained more than 10000 pictures in {\it R} filter (Bessell {\it R}) with exposures from 20 up to 30 seconds. Typical time of
unbroken monitoring of V0873 Per was 5-7 hours per night.
Calibration frames of base (bias) and dark frames (dark) with
required exposures were made in necessary numbers during every
night before and after observations of V0873 Per. Frames of flat
field (flat) were obtained by photographing of twilight sky.

Photometric data processing was made by aperture photometry method
using program
C-Munipack\footnote{http://c-munipack.sourceforge.net}. Optimum
aperture value corresponds to minimum value of standard deviation
for differential magnitudes. Deviations were calculated using reference stars 2MASS J02471801+4123073 and 2MASS J02470840+4120496 for different sets of observations. These stars assumed to be constant during observations.

Aperture is constant for all points during one night. Aperture
differences from night to night were insignificant. Precision
value for each observational point (single exposure) was in
the range $0.002^m-0.011^m$ for different nights.

Standard dark and flat field correction was done. Original data
files in fits-format were converted to text files containing
positions (JD) and research star brightness with respect to
non-variable check star on each source frame. Light curves of V0873 Per were created on the basis of these data.

\section{Light equation}

\begin{table}
\centering
\begin{minipage}{80mm}
\caption{Moments of light minima for V0873 Per from literature. ``I'' is a primary minimum, ``II'' is a secondary minimum. References in the table are the next: [1] \citet{samec2009}, [2] \citet{diethelm2010}, [3] \citet{diethelm2011}, [4] \citet{diethelm2012}, [5] \citet{kriwattanawong2015}, [6] Brno project, [7] \citet{hasanzadeh2013}, [8] \citet{diethelm2013}, [9] \citet{nagai2013}, [10] \citet{nagai2014}. O-C was calculated using ephemeris (\ref{eph3}) and (\ref{eph4}). With ``$^*$'' we mark moments of minima which are not shown in Figures \ref{f4} and \ref{f5} (see text for details).}
\label{lit}
\begin{tabular}{@{}cccc@{}}
\hline
HJD-2400000 & Min & Reference & O-C \\
\hline
54438.7605 & I & [1] & 0.0008 \\
54440.5298 & I & [1] & 0.0007 \\
54455.7199$^*$ & II & [1] & 0.0033 \\
54462.6464 & I & [1] & -0.0001 \\
54462.7943 & II & [1] & 0.0001 \\
54516.6131 & I & [1] & -0.0006 \\
55197.6874$^*$ & II & [2] & -0.0016 \\
55484.9228 & II & [3] & -0.0003 \\
55845.8839 & II & [4] & 0.0012 \\
55893.0680 & II & [5] & 0.0011 \\
55893.2152 & I & [5] & 0.0008 \\
55894.1001 & I & [5] & 0.0010 \\
55894.2475 & II & [5] & 0.0010 \\
55895.1321 & II & [5] & 0.0008 \\
55896.1646 & I & [5] & 0.0012 \\
56154.4989 &  I  & [6] & 0.0017 \\
56190.3242$^*$ & II & [7] & -0.0037 \\
56190.4764 & I & [7] & 0.0012 \\
56215.8368 & I & [8] & 0.0000 \\
56215.9849 & II & [8] & 0.0007 \\
56241.0505 & II & [9] & -0.0003 \\
56241.1995 & I & [9] & 0.0012 \\
56241.9360 & II & [9] & 0.0005 \\
56242.0841 & I & [9] & 0.0011 \\
56242.2304 & II & [9] & 0.0000 \\
56554.0883 & I & [10] & -0.0006 \\
56554.2359 & II & [10] & -0.0004 \\
\hline
\end{tabular}
\end{minipage}
\end{table}

\begin{table}
\centering
\begin{minipage}{80mm}
\caption{Moments of light minima for V0873 Per calculated from our observations. Observations in Crimea marked with ``Cr'', observation in Maidanak marked with ``M''.}
\label{obs}
\begin{tabular}{@{}cccc@{}}
\hline
HJD-2400000 & Min & Obs & O-C \\
\hline
56545.3896 & II & Cr & 0.0003 \\
56545.5370 &  I & Cr & 0.0002 \\
56548.4862 &  I & Cr & 0.0005 \\
56549.3711 &  I & Cr & 0.0006 \\
56549.5185 & II & Cr & 0.0006 \\
56568.2445 &  I & M & 0.0003 \\
56571.3407 & II & M & 0.0000 \\
56571.4884 &  I  & M & 0.0003 \\
56574.2898 & II & M & 0.0002 \\
56574.4372 &  I  & M & 0.0001 \\
56578.2712 &  I  & M & 0.0004 \\
56581.3669 & II & M & -0.0004 \\
56591.2463 &  I  & Cr & -0.0002 \\
56591.3938 & II & Cr & -0.0002 \\
56593.3102 &  I  & Cr & -0.0006 \\
56593.4576 & II & Cr & -0.0007 \\
56594.3427 & II & Cr & -0.0002 \\
56594.4902 &  I  & Cr & -0.0002 \\
56596.2594 &  I  & Cr & -0.0004 \\
56596.4069 & II & Cr & -0.0004 \\
56597.2917 & II & Cr & -0.0003 \\
56597.4391 &  I  & Cr & -0.0003 \\
56598.3236 &  I  & Cr & -0.0005 \\
56890.4236 & II & M & -0.0006 \\
56891.4560 &  I  & M & -0.0003 \\
56893.3725 & II & M & -0.0006 \\
56893.5202 &  I  & M & -0.0004 \\
56894.4046 &  I  & M & -0.0007 \\
\hline
\end{tabular}
\end{minipage}
\end{table}

To test V0873 Per on the existence of a third
body we collected most of published times of its light minima (see Table \ref{lit}\footnote{We do not use moment of minimum 2451370.8753 \citep{nicholson2006} following \citet{kriwattanawong2015}.}).
Times of light minima obtained from our own observations are shown in Table \ref{obs}.
These minima moments were determined in free search by solving
the light curves at fixed geometrical parameters
taken from previous study \citep{kriwattanawong2015} using a program\footnote{See \citep{kozyreva2001} for more details. Very similar program was described and used by \citet{khaliullina1984}. } designed to calculate orbital elements. To define moments of minima with the highest possible precision we use only full light curves with both branches as shown in Figure \ref{f1}. Our analysis of the data
in Tables \ref{lit} and \ref{obs} yields the following linear ephemerides to
predict primary (Min I) and secondary (Min II) minima:

\begin{equation}
\label{eph3}
Min\ I =HJD 2455892.9195 + 0.2949016\times E;
\end{equation}

\begin{equation}
\label{eph4}
Min\ II =HJD 2455893.0670 + 0.2949016\times E;
\end{equation}

\noindent Our value of binary orbital period $0.2949016$ days lies very close to value of period in Equation (3) from \citep{kriwattanawong2015}. Here $E$ is the number of orbital cycles since initial epoch.

Orbital parameters of a multiple system can be determined from its light equation\footnote{See Equation 3 in \citep{kozyreva2005}, also see \citet{martynov1948}. \citet{kozyreva1999} and \citet{kozyreva2005} reported about discoveries of third bodies in AS Cam and HP Aur binaries correspondingly using this method. \citet{lacy2014} independently confirm existence of a third body in HP Aur system.}. Solving the light equation by an iterative method of differential corrections we obtain following orbital parameters for the third body: orbital period $297\pm 15$ days, amplitude of light equation $85\pm 10$ seconds, orbital eccentricity of the third body $e_3=0\div 0.05$, $a\cdot \sin i = 0.17\pm 0.02$ astronomical units, here $a$ is the semi-major axis of the orbit of V0873 Per binary star's center of mass around the center of mass of the system ``binary star+third body'', $i$ is the angle between third body's orbital plane and the plane of the sky.
 
\begin{figure}
\includegraphics{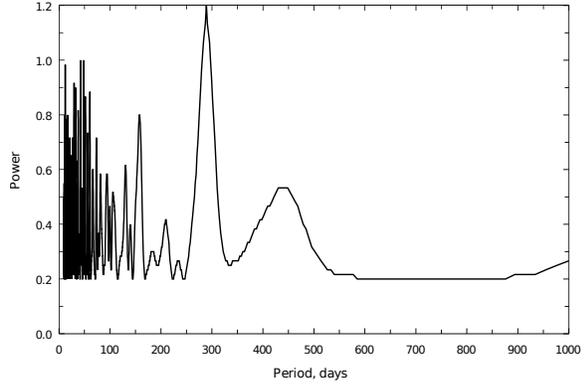}
\vspace{0pt} \caption{Power spetrum calculated using ephemeris (\ref{eph3}) and (\ref{eph4}). }\label{f2}
\end{figure}

\begin{figure}
\includegraphics{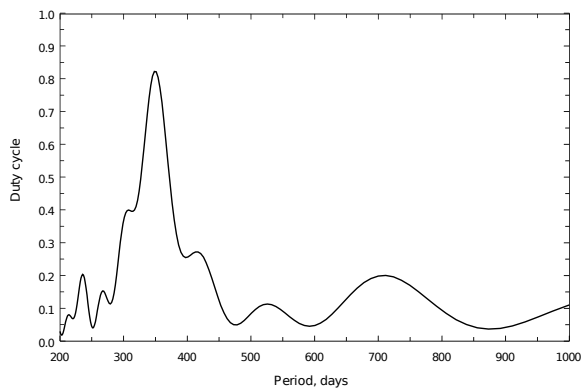}
\vspace{0pt} \caption{The dependency of a duty cycle from a signal period.}\label{f3}
\end{figure}

We conducted a Fourier analysis of data in assumption of zero eccentricity to study other possible periods which can be a solution of obtained moments of light minima using PERDET program \citep{breger1990}. As we can see from Figures \ref{f2} and \ref{f3} periods longer than $\approx 300^d$ have much less power. Only $\approx 40^d$ period has comparable power to our main period ($\approx 300^d$), but we have homogeneous period of observations during the same time interval, and we do not see any variations with such periods. Therefore the most suitable period is $\approx 300^d$.

We estimate value

\begin{equation}
\label{theta}
\theta=\frac{\sigma^2}{\sigma_0^2},
\end{equation}

\noindent where $\sigma$ and $\sigma_0$ are standard deviations calculated using formula

\begin{equation}
\label{sigma}
\sigma=\sqrt{\frac{1}{N-1}\Sigma (O-C)^2},
\end{equation}

\begin{figure}
\includegraphics{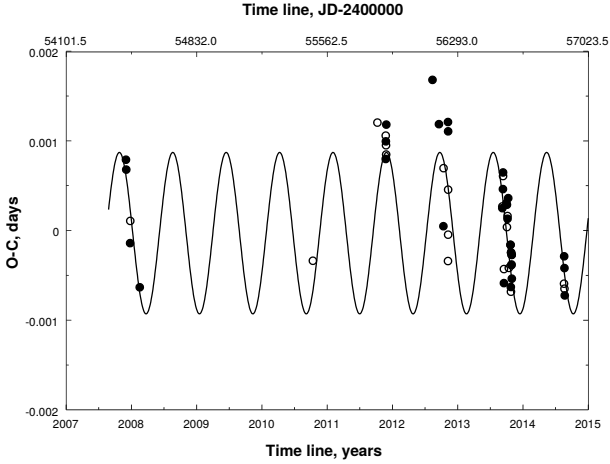}
\vspace{0pt} \caption{Light equation curve. We use ephemeris (\ref{eph3}) and (\ref{eph4}). Period and amplitude of light equation are $\approx 300$ days and 75 seconds respectively. Continuous line is the theoretical curve, while circles depict observational points. Filled circles correspond to primary minima I, open circles correspond to secondary minima II. See Tables \ref{lit} and \ref{obs} for details.}\label{f4}
\end{figure}

\begin{figure}
\includegraphics{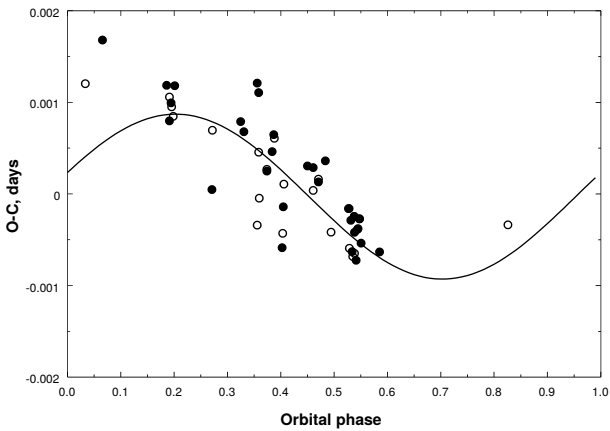}
\vspace{0pt} \caption{The same as in Figure \ref{f4}, for one orbital period of a third body.}\label{f5}
\end{figure}

\noindent In Equation (\ref{theta}) $\sigma _0$ corresponds to values from Tables \ref{lit} and \ref{obs}, while $\sigma$ was corrected with theoretical curve from Figures \ref{f4} and \ref{f5} arising from possible influence of a third companion. Values of $\sigma$ can be found in the Table \ref{tsigma}. Values of $\theta$ can be found in the Table \ref{ttheta}.

\begin{table}
\centering
\begin{minipage}{80mm}
\caption{Values of $\sigma$. $P$ is a corresponding periodical variation, column ``1'' presents corrected values, calculated using all moments of minima from Tables \ref{lit} and \ref{obs} (in this case $\sigma_0=0.00095$), while column ``2'' presents the same without points marked by ``*'' (in this case $\sigma_0=0.00079$). }
\label{tsigma}
\begin{tabular}{@{}ccc@{}}
\hline
$P$ & 1 & 2 \\
\hline
$\approx 300$ days & 0.00082 & 0.00056 \\
$\approx 4$ years & 0.00086 & 0.00062 \\
\hline
\end{tabular}
\end{minipage}
\end{table}

\begin{table}
\centering
\begin{minipage}{80mm}
\caption{The same as Table \ref{tsigma} for values of $\theta$.}
\label{ttheta}
\begin{tabular}{@{}ccc@{}}
\hline
$P$ & 1 & 2 \\
\hline
$\approx 300$ days & 0.74 & 0.50 \\
$\approx 4$ years & 0.83 & 0.62 \\
\hline
\end{tabular}
\end{minipage}
\end{table}

If we do not take into account three points \footnote{Because of their mean square-error is greater than 2.5$\sigma$. We can not offer an explanation of this high deviation and suppose that it arises from data processing and algorithms of minima time finding (for example, if individual observations of light curve during those nights were performed with significantly longer breaks between points in comparison with other observations) rather than from physical processes. } marked by ``$^*$'' in the Table \ref{lit}, we obtain $\theta=0.50$ for our period $\approx 300^d$ and $\theta=0.62$ for period $\approx 4$ years found by \citet{kriwattanawong2015}. This value is less than previous values for both periods of minima time variations and allows us to conclude that the system has light equation which we explain by the existence of a third body. In this case (without three marked points) orbital period of the third body is $\approx 300^d$ and amplitude of light equation is $\approx 85$ seconds.

We can not exclude 4 years variations found by \citet{kriwattanawong2015}. This fact does not allow to prefer one of these periods of time of minima variations and it is possible that both of them are valid, also there is a possibility that only one of these periodical variations is real.

\citet{kriwattanawong2015} could not make a final conclusion about nature of the time variation and choose between the magnetic activity and the gravitational influence of a third body on the period of V0873 Per. Therefore, future photometric and especially spectroscopic investigations of this system are required. Further text describes our result in terms of the existence of a third body.

Figure \ref{f4} shows residuals of observed times of minima (O) in Tables \ref{lit} and \ref{obs} from the theoretical moments of minima (C) calculated using ephemeris (\ref{eph3}) and (\ref{eph4}) depending on date of observations. Figure \ref{f5} shows the same for one orbital period of the third body, so (O-C) in this figure depends on its orbital phase.

\section{Mass of the new body}

In general task of precise determination of mass of a circumbinary third body is a separate complicated investigation. Inclination angle of the orbit of the third companion from our photometry study stays unknown, therefore at this paper we make only a simple estimation of low limit of the mass of the third body. This section is valid only if time variations of minima of V0873 Per caused by the presence of a third body.

Since orbital period of V0873 Per is considerably longer\footnote{Orbital period of the new body is $\approx 300^d$, while orbital period of V0873 Per is less than $0.3^d$. At the same time a third body is able to possess almost stable orbit if its orbital period begins from $\approx 3$ days, see Table 6 in \citep{holman1999}. So, one can apply Kepler's third law for such body (for rough estimations, because strong disturbances of the orbit should exist). } than orbital period of central binary star we can use Kepler's third law for our simple estimations of mass.

\begin{equation}
\label{kepler}
P_{orb}=0.1\frac{a^{3/2}}{M^{1/2}};
\end{equation}
 
\noindent where $P_{orb}$ is the orbital period of a third body in days, $a$ is the semi-major axis of its orbit in solar radii, $M$ is total mass of all companions of the system in solar masses. According to spectral type K0V of the primary component of V0873 Per and mass ratio 2.6 (by \citet{samec2009}) summary mass of central binary stars is $\approx 1 M_{\odot}$. Orbital period derived from our observations and information from literature is $\approx 300$ days. These quantities with equation \ref{kepler} give semi-major axis of the third body to be $\approx 208 R_{\odot}$. Amplitude of light equation is 85 seconds. During this interval of time light travels a half of the distance which the third body displaces the central binary. Consequently semi-major axis of the orbit of the center of mass of binary star V0873 Per around the center of mass of the system ``V0873 Per + third body'' is $\approx 37 R_{\odot}$. This gives us ratio of mass of the third body to mass of the central binary to be $\approx 0.18$, so the mass of the third body is $\approx 0.2 M_{\odot}$ and corresponds to a red dwarf. If the orbit of the third body is inclined with respect to the orbit of the central binary, an estimation of its mass should be increased.

The same estimation can be made for 4 years variations of times of minima of the binary. If this variation exists and the explanation of it is the gravitational influence of a third body, the mass of those body is $\approx 0.06 M_{\odot}$.

\section{Conclusions}

We report about discovery of a third body candidate in eclipsing variable V0873
Per. We found the periodical variation of orbital period of V0873 Per with a period $\approx 300$ days. This variation explains by the gravitational influence of a third companion on the central
binary star. The low limit of mass of this body is $\approx 0.2 M_{\odot}$. Moments of minima calculated from our observations can be used in future investigations of V0873 Per.

Investigations of multiplicity of stars play very important role in understanding of star formation processes (see e.g. \citet{tokovinin1999}). According to \citet{tokovinin2006} the fraction of solar type spectroscopic binaries reach up to 96\% for binaries with orbital period less than 3 days. Our investigation is in a good agreement with their statement.

V0873 Per is a very interesting object and requires additional investigations to verify variations of times of minima of the binary light curve with period of 300 days in this study and 4 year period found by \citet{kriwattanawong2015} to explain nature of this system (gravitational influence of additional bodies in the system and/or \citet{applegate1992} magnetic mechanism). Also a more detailed future study of V0873 Per probably can help to comprehensively clarify the influence of additional bodies on the central binary and dynamical evolution of orbits in triple/multiple systems, see e. g. some recent papers \citep{naoz2013,li2014}.

\section*{Acknowledgments}

We thank S. Yu. Shugarov, B. M. Shustov and A. M. Tatarnikov for
their kind help in building our team. We are grateful to anonymous referees for very useful comments.

\label{lastpage}

\end{document}